\UseRawInputEncoding
\documentclass[prl,twocolumn,floatfix]{revtex4}

\usepackage{graphicx}
\usepackage{amsmath}
\usepackage{xcolor}
\usepackage{ulem}
\usepackage[T1]{fontenc}
\usepackage{float}
\usepackage{comment}
\usepackage{physics}

\newcommand{\DS}[1]{\textcolor{red}{#1}}
\newcommand{\DSs}[1]{\textcolor{red}{\sout{#1}}}

\newcommand{\dif}{\mathrm{d}\;\!\!}

\newcommand{\Eblue}{E}
\newcommand{\Epurple}{\varepsilon}

\begin{document}

\preprint{APS/123-QED}

\title{Formation of Matter-Wave Polaritons in an Optical Lattice}

\author{Joonhyuk Kwon}
\author{Youngshin Kim}
\author{Alfonso Lanuza}
\author{Dominik Schneble}
\affiliation{Department of Physics and Astronomy, Stony Brook University, Stony Brook, NY 11794-3800, USA}

\date{\today}

\begin{abstract}
The polariton, a quasiparticle formed by strong coupling of a photon to a matter excitation, is a fundamental ingredient of emergent photonic quantum systems ranging from semiconductor nanophotonics to circuit quantum electrodynamics. 
Exploiting the interaction between polaritons has led to the realization of superfluids of light as well as of strongly correlated phases in the microwave domain, with similar efforts underway for microcavity exciton-polaritons.
Here, we develop an ultracold-atom analogue of an exciton-polariton system in which interacting polaritonic phases can be studied with full tunability and without dissipation. In our optical-lattice system, the exciton is replaced by an atomic excitation, while an atomic matter wave is substituted for the photon under a strong dynamical coupling. We access the band structure of the matter-wave polariton spectroscopically by coupling the upper and lower polariton branches, and explore polaritonic many-body transport in the superfluid and Mott-insulating regimes, finding quantitative agreement with our theoretical expectations. Our work opens up novel possibilities for studies of polaritonic quantum matter.

\end{abstract}
\maketitle

Since their first description as a superposition of light and matter excitations \cite{Hopfield1958,Pekar1958}, polaritons have been an essential key for understanding the nature of strong light-matter couplings in solids. While the polariton concept \cite{Mills1974} has been broadly expanded to various fields and contexts \cite{Basov2021}, its essence is captured by the microcavity polariton, whose realization in a semiconductor \cite{Weisbuch1992} first revealed exciton-photon mode splitting as the signature of a mobile, dynamically-coupled quasi-particle with a dispersion relation that hybridizes those of its two constituents.

Polariton interactions, mediated by the heavier constituent, open up novel possibilities for engineering effective photon-photon interactions that can lead to Bose-Einstein condensation \cite{Carusotto2013,Deng2010} and strongly interacting many-body states \cite{Ma2019,Carusotto2020}.
In this context, several polariton platforms including semiconductor microcavities \cite{Schneider2016,Cuevas2018}, Rydberg polaritons \cite{Carusotto2013}, as well as waveguide \cite{Chang2018} and circuit \cite{Carusotto2020} quantum electrodynamics provide breakthrough possibilities for quantum simulations \cite{Hartmann2016, Noh2016}.

In polaritonic systems, it is an important challenge to achieve strong coupling \cite{Liu2017} and interactions \cite{Sundaresan2019} with tunable parameters, while managing dissipation \cite{Ma2019}; in addition, some systems can be impeded by imperfections that limit their scalability \cite{Blais2020}. To study fundamental properties of polaritons in the context of many-body quantum simulation, a platform that avoids the aforementioned complications would therefore be desirable.

In this Letter, using a recently developed experimental approach \cite{Krinner2018,Stewart2020}, we introduce a polariton platform featuring full flexibility and no intrinsic losses, in which the photonic constituent is replaced by a matter wave \cite{GTudela2018C}, thus forming matter-wave polaritons.
While preserving the fundamental features of conventional polaritons~\cite{Basov2021}, the ratio of effective mass and interactions of these novel polaritons is fully tunable, and the system is dissipation-free with an infinite Purcell factor. By controlling the parameters in the regime of interest, our system can simulate fundamental polaritonic properties in the regime of strong interactions \cite{Hartmann2006,Greentree2006,Shi2018}.

In analogy to an exciton-polariton, we realize a quasiparticle that hybridizes the quadratic dispersion relations of two constituents of disparate mass. While the light constituent playing the role of the photon in the microcavity is a free-space atomic matter wave, the other constituent, replacing the exciton, is an atom with high effective mass induced by an optical lattice, that is coupled to the matter wave via an effective dipole moment. The coupling hybridizes the two dispersion relations into lower (LP) and upper (UP) polariton branches. In the limit of a strongly-localized heavier constituent, the bound state reported in \cite{Krinner2018} can be understood as a partial feature of the LP branch.

The optical lattice confining the atom not only increases its effective mass but also provides an analog of strongly-correlated polaritonic arrays \cite{Schneider2016}, in which the interaction between excitons is replaced by the natural collisional interaction between the heavy atoms in the lattice. We create a system of strongly-interacting polaritons, in which the competition between hopping and interactions lead to a tunable phase transition from a polaritonic superfluid to a Mott insulator, in agreement with expectations based on the polariton band structure \cite{Lanuza2021,Shi2018}.

\begin{figure}[t!]
\centering
    \includegraphics[width=.95\columnwidth]{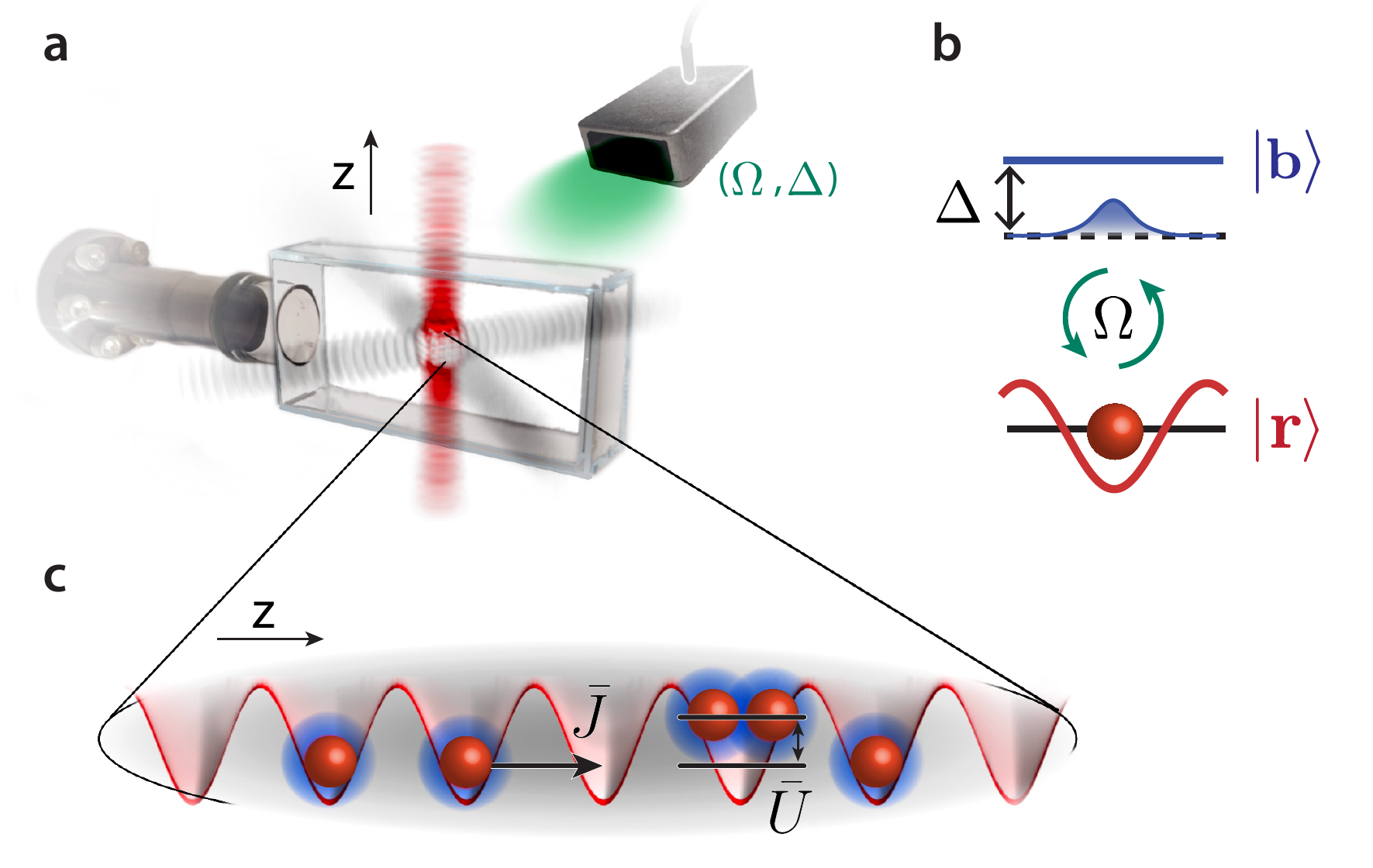}
    \caption{\textbf{Experimental scheme and polariton formation}. \textbf{a,b,} An $\ket{r}$ (red) atom in the ground band of a deep state-selective optical lattice (red) is coupled through a microwave field (green) with strength $\Omega$ and detuning $\Delta<0$ to an unconfined state $\ket{b}$ (blue) with energy below the edge of the continuum of motional states. The atomic ensemble is confined to a system of tightly confining tubes created with state-independent transverse lattices (gray). \textbf{c,} In each tube, the coupling results in a polaritonic superposition of lattice-trapped atoms and matter-wave modes with quasi-particle hopping $\bar{J}$ and on-site interaction $\bar{U}$. 
    }
\label{FIG:SystemIntro}
\end{figure}

The scheme for our experiments is illustrated in Fig.~\ref{FIG:SystemIntro}. Using an optically-trapped Bose-Einstein condensate of around $10^4$ $^{87}$Rb atoms in the hyperfine ground state $\ket{r}\equiv\ket{F=1, m_F=-1}$, we load an array of $10^3$ red-detuned, tightly-confining optical lattice tubes with depth $s_\perp\gg1$ measured in terms of the recoil energy $E_{r\perp}=h^2/2m\lambda_\perp^2$, where $\lambda_\perp=1064$~nm is the lattice laser wavelength and $m$ the atomic mass. Another lattice (``$z$-lattice'') with variable depth $s$, in terms of a corresponding recoil energy $E_r=\hbar\omega_r$ at wavelength $\lambda$, additionally confines the $\ket{r}$ atoms along the vertical $z$ axis aligned with the tubes. The $\ket{r}$ atoms hop between sites of the $z$ lattice with finite tunnel coefficient $J$, while transport across tubes is negligible on experimentally relevant time scales. The atoms are coupled to a second hyperfine ground state, $\ket{b}\equiv\ket{F=2, m_F=0}$ via a 6.8 GHz microwave field of strength $\Omega$ and negative detuning $\Delta$ from the atomic resonance in the $z$-lattice, which is shifted by the difference in zero-point energy. The wavelength $\lambda=790.0$~nm and polarization ($\sigma^-$) of the $z$-lattice (for which $E_r=h\times 3.67~$kHz) are chosen such that it is fully state-selective and $\ket{b}$ atoms do not experience its potential at all, and can freely move along the tubes (for times smaller than $2\pi/\omega_z \sim 10$~ms) occupying a continuum of modes $\ket{p}$ with momentum $p$.  

As observed in \cite{Krinner2018}, for the radiative decay of an atom from a lattice well, the coupling to the free modes for $\Delta<0$ induces the formation of a bound state containing an evanescent $\ket{b}$ matter wave with decay length $\sim \sqrt{\hbar/2m|\Delta|}$.
If now the evanescent wave starts to leak into neighboring sites, the coupling between the two components can induce an effective tunneling of the $\ket{r}$ atom, and since the evanescent wave remains localized around the atom, this process then corresponds the hopping of a quasiparticle. In a many-body system, such matter-wave polaritons will be characterized by an effective tunneling matrix element $\bar{J}$ and an onsite interaction $\bar{U}$. Here we explore the signatures of polariton formation in a Bose-Hubbard scenario in which the overall state of the system is tuned via the ratio between tunneling and onsite interactions.

\begin{figure}[t!]
\centering
    \includegraphics[width=1.0\columnwidth]{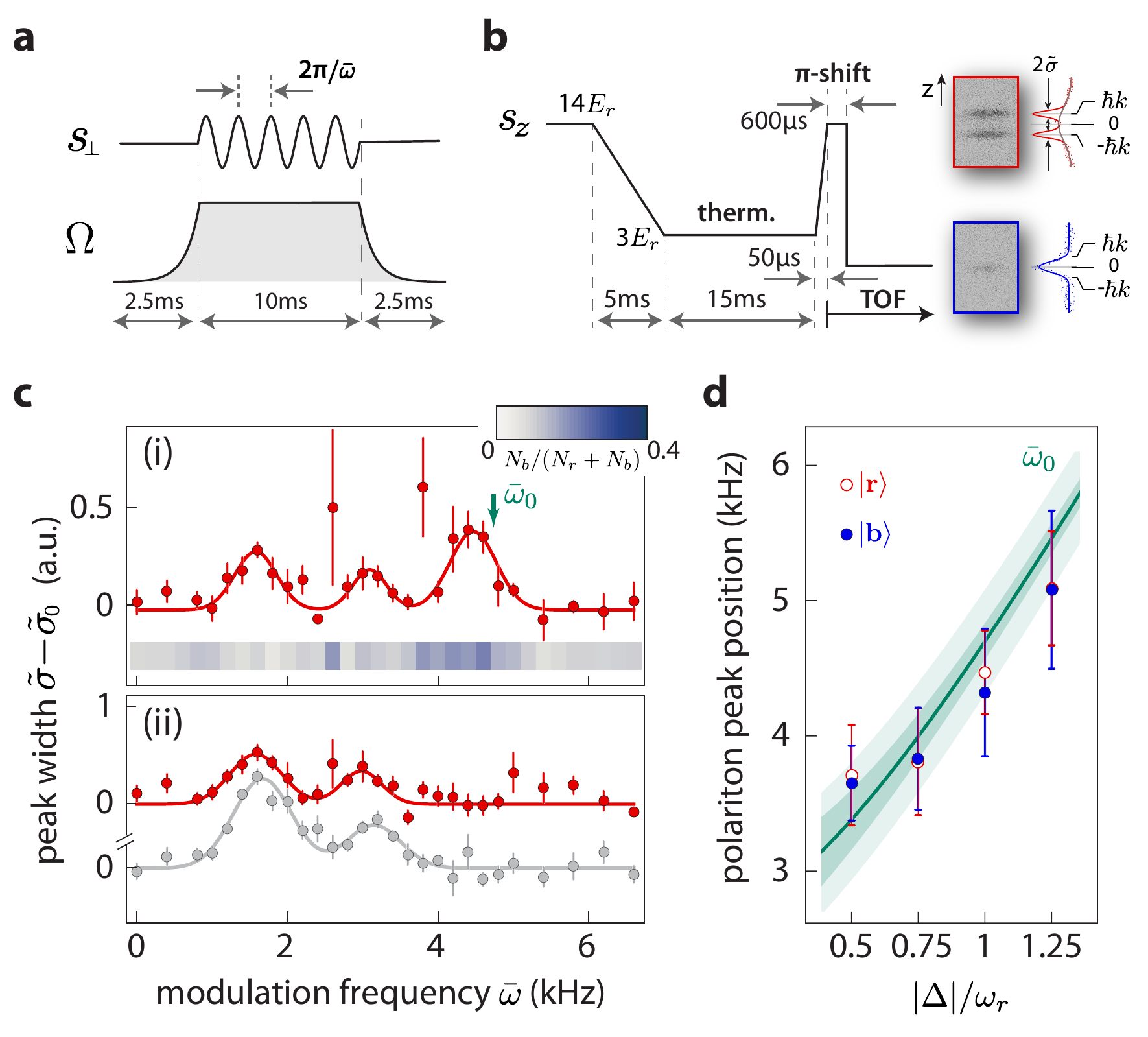}
    \caption{\textbf{Excitation spectra in the Mott regime} ($s_\perp=40$, $s_z=14$).
    \textbf {a,} Excitation: after ramping up $\Omega$, $s_\perp$ is modulated by $\pm30\%$ at variable frequency $\bar{\omega}$ for 10~ms. \textbf{b,} Detection sequence involving rethermalization and a gravitational $\pi$-phase shift for the $\ket{r}$ atoms after the coupling is turned off (details see methods). 
    The two panels display the  $\ket{r}$ and $\ket{b}$ momentum distributions measured after resonant excitation at detuning $\Delta/\omega_r=-1.00(7)$ and coupling $\Omega/\omega_r=1.09(2)$. 
    The width $\tilde{\sigma}$ is extracted through Gaussian fits to the 1D density along $z$. 
    \textbf{c,} Excitation spectra (i) for $\Delta,\Omega$ as in \textbf{b}, with the inset stripe showing the number of $\ket{b}$ atoms; (ii) for the uncoupled case $\Omega=0$ (red) and additionally $s_z$=10 (gray). \textbf{d,} Detuning dependence of the third excitation feature. Open red circles denote the $\ket{r}$ peak position, and blue dots denote the position of the maximum transfer of $\ket{b}$ atoms, as extracted from Gaussian fits, with error bars giving the spectral width of the excitation. The green curve is the calculated polariton excitation energy $\bar{\omega}(\Delta,\Omega)$, where the shaded areas include the width of the ground band (darker green) and the uncertainties in $\Delta$ ($\pm 0.27$~kHz) and $\Omega$ ($\pm2\%$) (lighter green).}
\label{FIG:onsite}
\end{figure}

To explore the effects of the vacuum coupling, we first measure the excitation spectrum deep in the Mott regime with a small ratio $J/U\sim5\times10^{-3}$ between atomic tunneling and onsite interactions. The procedure is summarized in Fig.~\ref{FIG:onsite}\textbf{a} and \textbf{b}. After preparing the system at $s_{z}=14$ and $s_{\perp}=40$ (where $U = h\times 1.7$~kHz), we exponentially ramp up $\Omega$ at fixed $\Delta$ on a time scale $\tau\gg1/|\Delta|$ that is long enough to preclude nonadiabatic shedding of matter waves \cite{Krinner2018}. With the coupling $\Omega$ applied, we sinusoidally modulate $s_\perp$ at variable frequency $\bar{\omega}$ in order to induce resonant excitations of the gapped Mott phase \cite{Stoeferle2004} (we note that, while varying $s_\perp$ changes the polariton onsite interaction $\bar{U}$, it does not directly affect the single-polariton feature $\Delta$). After ramping $\Omega$ back down, we enter the superfluid regime in which we characterize the effects of the modulation, which are in the form of a reduction of coherence, via time-of-flight (ToF) measurements. 

The effects of the coupling on the excitation spectrum are shown in  Fig.~\ref{FIG:onsite}\textbf{c}. In a reference run without coupling, we observe excitation peaks near $\bar{\omega}=U/\hbar$ and $2U/\hbar$, as expected from resonant atom redistribution between sites \cite{Kollath2006}. The coupling has no noticeable effect on the position of the peaks, suggesting that for the parameters used ($\Omega,|\Delta|\sim\omega_r$) the change of the on-site energy is small. However, an additional feature appears, centered at a larger modulation frequency $\bar{\omega}$ not far from $|\Delta|$ (see Fig.~\ref{FIG:onsite}\textbf{d}), accompanied by excess $\ket{b}$ population left after the coupling is ramped back down. 

In order to discuss these observations further, we first develop a quantitative description of the expected polariton features, following the general approach of \cite{Shi2018}. We first consider a $\ket{r}$ atom coherently distributed over the sites of the $z$-lattice, such that the system Hamiltonian takes the form of a series of coupled Weisskopf-Wigner models. By expressing the momentum of the vacuum modes $p_{n,q}$ in terms of a quasimomentum $q\in [-k,k]$ and an integer band index $n\geq1$, the non-interacting part of the Hamiltonian decouples as $\hat{H} = \sum_{n,q}\varepsilon_{n,q} \hat{c}^\dagger_{n,q}\hat{c}_{n,q}$, where $\hat{c}^\dagger_{n,q} = \alpha_{n,q} \hat{r}_q^\dagger + \sum_{n\prime}\beta_{n,n\prime,q}\hat{b}^\dagger_{n^\prime,q}$ creates a matter-wave polariton as a $(\Omega,\Delta)$-dependent superposition of a $\ket{r}$ Bloch wave and all $\ket{b}$ modes of the same $q$. The corresponding polariton dispersion relation $\varepsilon_{n,q}$ emerges from the equation $\varepsilon_{n,q} - \hbar\Delta_q = \sum_{n^\prime}\hbar^2g_{n^\prime,q}^2 /\left(\varepsilon_{n,q}-E_{n^\prime, q}\right)$ (see  methods and \cite{Lanuza2021}), where $g_{n,q} = \Omega \braket{\phi_0}{n,q}/2$ is the coupling strength between the originating lattice Wannier function $\ket{\phi_0}$ and the free-momentum mode $\ket{n,q}$ normalized to a Wigner-Seitz cell, $E_{n,q} = p_{n,q}^2/2m$ is the free-atom dispersion, and $\Delta_q = \Delta + 4 J \sin^2(\pi q/2k)$ a $q$-dependent detuning from the continuum edge that accounts for the bandwidth $4J$ of the $z$-lattice (for details see appendix). The $n=1$ band is the LP branch of the matter-wave polariton, while the bands with $n\geq2$ produce a series of UP branches.

\begin{figure}[t!]
\centering
    \includegraphics[width=\columnwidth]{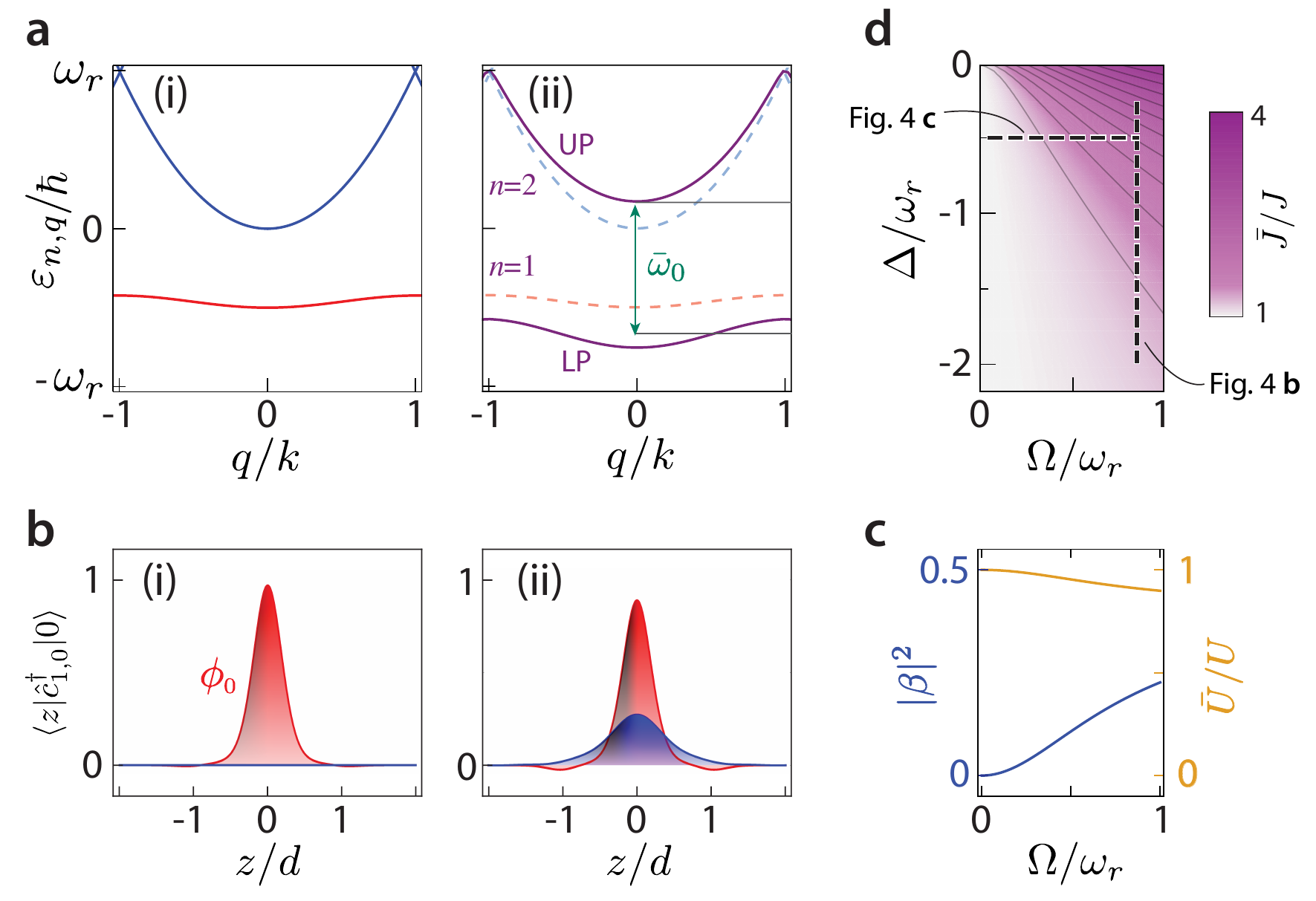}
    \caption{\textbf{Polariton band structure}, calculated for $s_z=10$. \textbf{a,} (i) Uncoupled band structure for lattice-trapped atoms in $\ket{r}$ (red curve) and free atoms $\ket{b}$ (blue curve). (ii) Band structure for a coupling $\Omega/\omega_r=1$ and a detuning $\Delta/\omega_r=-0.5$, with the two lowest polariton bands ($n=1,2$) shown as purple curves. The green arrow denotes the separation $\bar{\omega}_0$ from the middle of the lower band to the bottom of the upper band. The dashed curves reproduce the atomic band structure. \textbf{b,} Wannier functions obtained for the two scenarios of \textbf{a}, containing $\ket{r}$ (red) and $\ket{b}$ (blue) components, where $d=\lambda/2$.
    \textbf{c,} Fraction of the $\ket{b}$ component $\lvert\beta\rvert^2$ and polariton interaction energy $\bar{U}$ as a function of $\Omega$ for $\Delta/\omega_r=-0.5$. \textbf{d,} Polariton ground-band tunneling coefficient $\bar{J}$ relative to $J$ as a function of $\Omega$ and $\Delta$. The dashed lines correspond to measurements shown in Fig.~\ref{FIG:hopping}\textbf{b} and \textbf{c}.
    }
\label{FIG:theory}
\end{figure}

Fig.~\ref{FIG:theory}\textbf{a} depicts the ground (lower) and first excited (upper) polariton bands, $\varepsilon_{0,q}$ and $\varepsilon_{1,q}$ at a typical detuning. In real space, taking an atom localized in site $j$, the coupling induces, via $\hat{c}^\dagger_{n,j}= \sum_q \hat{c}_{n,q}^\dagger e^{-iqz_j}$, the formation of a matter-wave polariton as a localized quasi-particle. Its ground-band Wannier function $\mel{z}{\hat{c}^\dagger_{1,0}}{0}$ is shown in Fig.~\ref{FIG:theory}\textbf{b}. As a result of the coupling, the $\ket{r}$ component acquires additional amplitude in neighboring sites, which can be seen as being due to a re-coupling of the evanescent $\ket{b}$ tail into the lattice potential.

We estimate the effect of the coupling on the on-site interaction in the $z$-lattice by setting $\bar{U}\approx \bar{g}\int\norm*{\mel{z}{\hat{c}^\dagger_{1,0}}{0}}^4 dz$, where we take the one-dimensional collisional strength $\bar{g}$ as state-independent (for $^{87}$Rb, the differences are on the percent level), such that we capture the dominant effects of the modified spatial overlap between the components. The magnitude of the shift $\delta U=(U-\bar{U})$ depends on both the extent of the evanescent tail and its amplitude; results for our parameters are shown in Fig.~\ref{FIG:theory}\textbf{c}. Consistent with our observation, the maximum shift $\delta U/U\sim0.1$ is below what our spectroscopy method can systematically resolve, as is also evidenced in a reference measurement without coupling in which we correspondingly lowered $U$ by reducing the $z$-lattice depth (see Fig.~\ref{FIG:onsite}\textbf{c}).

Based on the polariton band structure, we now discuss the mechanism responsible for the resonant excitation of $\ket{r}$ atoms in the lattice that is accompanied by the appearance of $\ket{b}$ atoms. Polaritons in the lower band are mostly in the $\ket{r}$ state, which is tightly confined. As they are periodically squeezed in the orthogonal direction, they are subject to a strong perturbation of their interaction energy $\bar U$. If this is done resonantly ($\bar\omega\approx\bar\omega_0$), they can be excited into the upper band, as it is mostly composed of the unconfined $\ket{b}$ state and thus is less sensitive to such perturbations. The third peak in the excitation spectrum (see Fig.~\ref{FIG:onsite}\textbf{c}) located around the frequency difference $\bar{\omega}_0$ between the bottom of the UP band and the center of the LP band, $\hbar \bar{\omega}_0 =\varepsilon_{2,0}- \sum_q\varepsilon_{1,q}$ corresponds to a conversion from the deeply-bound Wannier polariton state to quasi-free states in the region where the density of states is highest. The band gap increases with $|\Delta|$, requiring higher modulation frequencies $\bar{\omega}$ to excite this resonance (c.f. Fig.~\ref{FIG:onsite}\textbf{d}). Because the UP is dominated by $\ket{b}$, the excitation from the LP leaves $\ket{r}$ vacancies in the $z$-lattice, and atoms are instead observed as excess $\ket{b}$ population around zero momentum, as seen in the inset in Fig.~\ref{FIG:onsite}\textbf{b}. 

\begin{figure}[t!]
\centering
    \includegraphics[width=1.0\columnwidth]{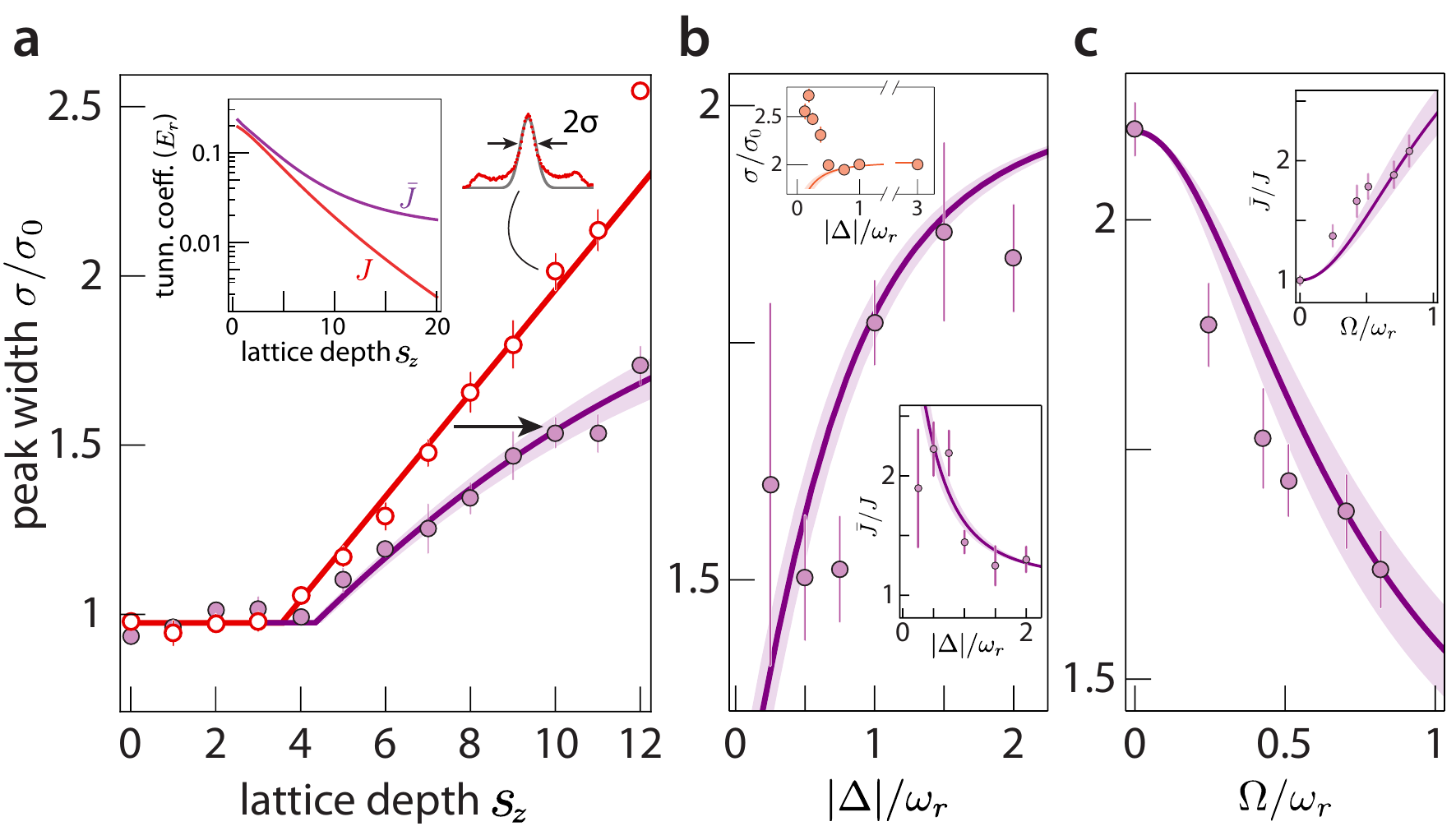}
    \caption{\textbf{Renormalization of hopping} extracted from the coherence of the $\ket{r}$ component at $s_\perp=18$.
   \textbf{a,} Scan over $s_z$ for $\Delta/\omega_r=-0.50(7)$,  $\Omega/\omega_r=0.84(2)$ (purple), compared to a reference measurement with $\Omega=0$ (red). Plotted is the fitted 1D Gaussian width $\sigma$ along $z$ of the central diffraction peak in time-of-flight (see inset). 
   The red curve is a piece-wise linear fit to the reference, with slope set to zero left of the kink, which is used to establish the relationship between $\sigma$ and $s_z$. The purple curve is obtained from the red curve after shifting $s_z(J)$ (black arrow) according to the expected rescaling of $J$ to $\bar{J}$; the shaded area accounts for the experimental uncertainties in $\Delta$ and $\Omega$. Inset: tunneling coefficients $J$ (red) and $\bar{J}$ (purple) as a function of $s_z$. \textbf{b, c,} Parameter scans at $s_z=10$ according to the traces shown in Fig.~\ref{FIG:theory}\textbf{d}, with variable $\Delta$ at $\Omega/\omega_r=0.84(2)$ (\textbf{b}), and with variable $\Omega$ at $\Delta/\omega_r=-0.50(7)$ (\textbf{c}). The theoretical prediction reflecting experimental uncertainties is shown as a solid purple line surrounded by shaded areas; the insets in the bottom half of \textbf{b} and in \textbf{c} show the corresponding ratios $\bar{J}/J$. The inset in the top half of \textbf{b} shows the peak width for $\Omega/\omega_r=0.28(1)$ (and $\hbar\Omega/U=1.01(1)$). Vertical error bars show the standard error of the mean; horizontal error bars are less than the size of the data points. All data points are the average over at least 3 runs. 
  }
\label{FIG:hopping}
\end{figure}

To further elucidate these features, we study the hopping in a polariton Bose-Hubbard model. Compared to the modification of the onsite interaction $\bar{U}$, the renormalization of $J$ for the atoms to $\bar{J} = -\sum_q\varepsilon_{1,q}e^{i \pi q/k}$ for the polaritons can be significant. In order to test the magnitude of this effect, we compare the degree of coherence of the $\ket{r}$ component at different lattice depths $s_z$ with that of the (coupling-free) atomic reference with known $J\equiv J(s_z)$. Using the width of the central peak of the $\ket{r}$ diffraction pattern as a proxy \cite{Kollath2004}, we record the ToF distribution after switching off the optical potential and the coupling and removing the $\ket{b}$ component with resonant light (to eliminate four-wave mixing \cite{Pertot2010}). The results for $\Delta/\omega_r=-0.5$ and $\Omega/\omega_r= 0.8$ are shown in Fig.~\ref{FIG:hopping}\textbf{a}. The peak width $\sigma$ of the reference is well fit by a piecewise-linear function with fixed zero slope in the superfluid region, and positive slope in the Mott regime \cite{Gadway2011}. When we turn on the coupling, the measured peak width $\bar{\sigma}$ is comparable in the superfluid region and then increases with $s_z$, but not as much as the reference.

We use the fitted reference curve and the functional dependence $J(s_z)$ obtained from the uncoupled band structure to obtain an operational relationship $\sigma(J)$ between peak width and hopping. Under the assumption that the onsite interaction remains unchanged, the same relationship should hold between $\bar{\sigma}$ and $\bar{J}$ obtained from the polariton band structure, and indeed we find that the rescaled reference curve quantitatively reproduces the observed peak widths for the experimental coupling parameters. In the $\ket{r}$ basis, the polariton formation effectively renormalizes the depth of the applied lattice, leading to a horizontal shift of points of a given peak width towards increasingly larger lattice depths. Near the kink at $s_z=4$, the rescaling of hopping implies both the presence of a superfluid-to-Mott transition of polaritons, and of a vacuum-coupling driven transition from the singly-occupied atomic Mott lobe to a polariton superfluid.

The dependence of the hopping on the coupling parameters $\Delta$ and $\Omega$ is shown in Fig.~\ref{FIG:hopping} (b, c). The measured peak width changes as expected from the rescaled reference curve, and the inferred polaritonic enhancement $\bar{J}/J$ of the hopping agrees well with the prediction of the polariton band structure. In particular, unlike $J$, $\bar{J}$ saturates for large $s_z$ as a result of matter-wave-induced hopping.

In Fig.~\ref{FIG:hopping}, there is good agreement between experiment and theory if at least one of the coupling parameters is comparable to $\omega_r$. However, for weaker couplings near the continuum edge, a reduction of $|\Delta|$ can lead to an increase of the peak width, as seen in the upper inset of Fig.~\ref{FIG:hopping}\textbf{b}, a behavior that cannot be explained by a violation of adiabaticity in applying the coupling ($1/\tau \sim 0.02~\omega_r$). For small $\Omega$ and $|\Delta|$, the $\ket{b}$ component of the LP has large spatial extent and low density, such that the collisional on-site interaction between the tightly confined $\ket{r}$ atoms can lead to an effective change of the detuning. Indeed, the divergence of the peak width occurs near $|\Delta|/\omega_r\sim U/E_r=0.28$, consistent with a shift to effectively positive detunings, at which we expect radiative matter-wave decay to give rise to a quantum Zeno effect \cite{Syassen2008}. We note that similar interaction-activated dissipation that inhibits transport in a Bose-Hubbard model has recently been observed using photon-induced losses \cite{Tomita2017}.

In this work, we focused on the lower polariton branch, corresponding to renormalized hopping of excitons in a material system \cite{Schneider2016}, which has allowed us to realize a polaritonic quantum phase transition as well as a transition from an atomic Mott insulator to a polariton superfluid. By exploiting the lattice band structure and applying specific couplings, exotic polariton band structures can be implemented \cite{Lanuza2021}, including those featuring frustration. We can also focus on the upper branch dominated by $\ket{b}$ atoms, which corresponds to the renormalized transport of photons.  Making the $z$-lattice state-dependent (instead of state-selective \cite{Stewart2020}) will allow us to create the equivalent of coupled-cavity arrays, and implement the analogue of photon blockade~\cite{Birnbaum2005} for $\ket{b}$ atoms, with the strongly interacting $\ket{r}$ state playing the role of a nonlinearity. Introducing couplings between more than two excitonic or photonic bands should enable studies of analogues of multiexciton polaritons \cite{Ouellet2015}, multimode strong coupling \cite{Sundaresan2015}, and spin-orbit coupling \cite{Solnyshkov2021}. This may open up possibilities for studying topological polaritonic systems in higher dimensions \cite{Karzig2015}. 

We thank M. Stewart, M. G. Cohen, Y. Li and T-C. Wei for discussions, and M.G.C. for a critical reading of the manuscript. This work was supported by NSF PHY-1912546, with additional funds from SUNY Center for Quantum Information Science on Long Island.

\bibliographystyle{apsrev}

\newpage

\subparagraph{\bf {\centerline{Supplementary Material}}}

\begin{figure}[ht!]
\centering
    \includegraphics[width=0.55\columnwidth]{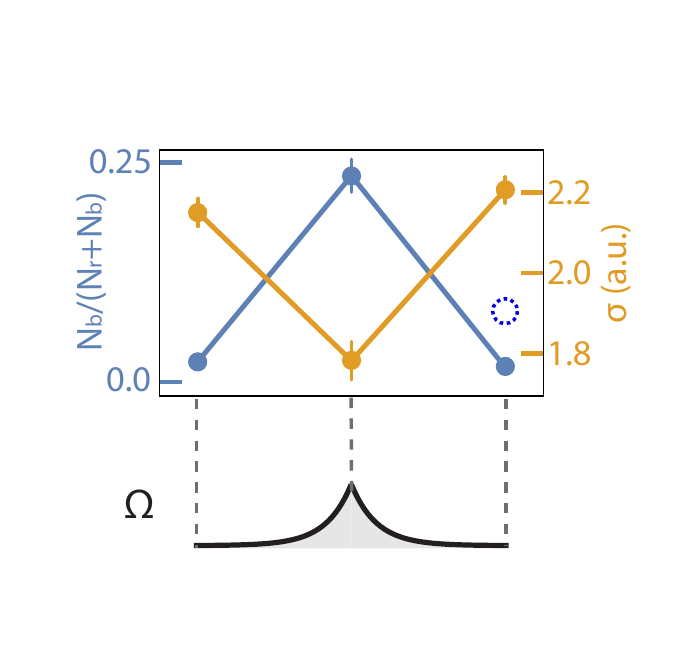}
    \caption{ Fraction of $\ket{b}$ atoms (blue) and peak width of $\ket{r}$ atoms (orange) after a symmetric microwave ramp up and down (each over 2.5~ms) for the parameters of Fig.~\ref{FIG:hopping} at $s_z=10$. The dotted open blue circle is the blue atom fraction for the sequence of Fig.~\ref{FIG:onsite}~\textbf{c} without lattice modulation applied. We suspect that the degradation comes from magnetic-field noise that is able to drive LP$\to$UP transitions by directly affecting $\Delta$.
  }
\label{FIG:sup_adiabaticity}
\end{figure}

\subparagraph{\it System preparation and detection.} Our experiments start with BECs of $1\times 10^4$ $\ket{r}$ atoms in an optical trap \cite{Krinner2018_2}. For the measurements of Fig.~\ref{FIG:onsite}, we first ramp up the tubes over $150$~ms to $s_\perp=40$ using an exponential ramp, followed by an exponential ramp of the $z$ lattice to $s_z=14$ over $25$~ms. Following with the reversibly-applied microwave coupling (see Fig.~\ref{FIG:sup_adiabaticity}) and lattice modulation spectroscopy, we detect excitations through changes in the coherence back in the superfluid phase \cite{Gadway2011_2}. We first linearly ramp down the $z$-lattice to $s_z = 3$ over $5$~ms followed by $15$~ms of thermalization. The tubes and the optical trap are then switched off and the $z$-lattice depth is jumped to $s_z=14$ (in $50$~$\mu$s) and held for $600$~$\mu$s to induce a gravitationally-induced $\pi$ shift between the wells. The resulting diffraction pattern of $\ket{r}$ atoms features two zeroth-order peaks separated by $2\hbar k$ in the $z$ direction. We sum over the transverse direction, fit the two peaks on top of a thermal background with three Gaussians, and extract the width $\bar{\sigma}$ as the average over the two peaks. We note that the momentum distribution of the $\ket{b}$ atoms remains unaffected by the changes in $s_z$, since resonant collisions during the rethermalization phase, as well as collinear four-wave mixing in time-of-flight \cite{Pertot2010_2}, are both suppressed due to the mismatch of the $\ket{r}$ and $\ket{b}$ dispersion relations \cite{Gadway2012}, and the strong confinement in the tubes limiting coherence, respectively.

For the measurements in Fig.~\ref{FIG:hopping}, we use a smaller tube depth, $s_\perp=18$, and ramp up the $z$ lattice more slowly, over $80$~ms (starting in the middle of the $s_\perp$ ramp) in order to maximize the system coherence over a large range of parameters. After switching off all potentials, we fit the 1D diffraction patterns using a single background-free Gaussian to extract the width $\sigma$ of the central peak after summing up over the transverse direction. To eliminate possible effects of collinear four-wave mixing in ToF expansion, the $\ket{b}$ atoms are removed using a short blast pulse of resonant cycling light \cite{Pertot2010_2}.

For atom detection, we use standard absorption imaging on the $F=2\rightarrow~F^\prime=3$ cycling transition after a ToF of $15$~ms, preceded by Stern-Gerlach separation of the $\ket{r}$ and $\ket{b}$ states. The $F=1$ atoms are detected after transferring them to $F=2$ using a short repump pulse.

\subparagraph{\it Resonance condition.} In our experiments, we need to accurately determine and maintain the resonance condition ($\Delta=0$) between the $\ket{r}$ in the $z$-lattice and the $\ket{b}$ mode continuum edge. We use lattice transfer spectroscopy \cite{Reeves2015} to adjust and monitor the resonance condition interleaved with a set of measurements, with an extrapolated drift of maximally $300$~Hz between shots. For this purpose, we prepare the BEC in the $\ket{b}$ state, and then apply $400 \mu s$ long Rabi pulses ($\Omega=2\pi \times 1.0$~kHz) to transfer a small fraction ($<30 \%$) of atoms into the $z$ lattice, with calculated mean-field shifts well below 100~Hz. 


\subparagraph{\it Polariton band structure.}

In the Schr\"{o}dinger picture, our system Hamiltonian takes the form \cite{Stewart2017}
\begin{equation}
\begin{aligned}
\hat H_S&=
\sum_j \hbar\omega^{(r)}_0 \hat{r}_j^\dagger\hat{r}_j+\sum_{n,q} \hbar\omega^{(b)}_{n,q} \hat{b}_{n,q}^\dagger \hat{b}_{n,q}
\\&
+\sum_j\sum_{n,q} \hbar g_{n,q}\left(e^{-i(\omega_\mu t+qz_j)} \hat{r}_j\hat{b}^\dagger_{n,q}+\text{H.c.}\right),
\end{aligned}
\end{equation}
where
$\hbar\omega^{(b)}_{n,q}=\hbar\omega^{(b)}_0+\Eblue_{n,q}$ is the energy (internal plus kinetic) of the $\ket{b}$ atoms, $\hbar\omega^{(r)}_0$ is the energy of the $\ket{r}$ atoms, $\omega_\mu$ is the frequency of the microwave which determines the detuning through $\Delta=\omega_\mu-(\omega^{(b)}_0-\omega^{(r)}_0)$,
 $z_j=j\pi/k$ is the position of the $j^\text{th}$ site,
$\sum_j\equiv\sum_{j=-\infty}^{+\infty}$ and $\sum_{n,q}\equiv\sum_{n=1}^\infty\int_{-k}^{+k}\frac{\dif q}{2k}$.
The external action of the microwave makes this Hamiltonian time-dependent. However, by writing $\hat H_S=\hat H_{0,S}+\hat H_{1,S}$ with
 \begin{equation}
\begin{aligned}
\hat H_{0,S}&=
\sum_j \hbar(\omega^{(b)}_0-\omega_\mu) \hat{r}_j^\dagger\hat{r}_j+\sum_{n,q} \hbar\omega^{(b)}_0 \hat{b}_{n,q}^\dagger \hat{b}_{n,q}
\end{aligned}
\end{equation}
one gets the time-independent interaction Hamiltonian $\hat H\equiv \hat H_{1,I}=e^{i\hat H_{0,S} t/\hbar}\hat H_{1,S}e^{-i\hat H_{0,S} t/\hbar}$
 \begin{equation}
\begin{aligned}
\hat H&=
\sum_j \hbar\Delta \hat{r}_j^\dagger\hat{r}_j+\sum_{n,q} \Eblue_{n,q} \hat{b}_{n,q}^\dagger \hat{b}_{n,q}
\\&
+\sum_j\sum_{n,q}\hbar g_{n,q}\left(e^{-iqz_j} \hat{r}_j\hat{b}^\dagger_{n,q}+\text{H.c.}\right).
\end{aligned}
\end{equation}

Following the approach of \cite{Shi2018_2} (see \cite{Lanuza2021_2} for an alternative), it is then possible to study the emergence of polaritons.  Fourier-transforming the operators
\begin{equation}\label{eq:operatorFT}
\hat r_q=\sum_j e^{-iqz_j}\hat r_j, \qquad \hat r_j=\sum_q e^{iqz_j} \hat r_q
\end{equation}
leads to
 \begin{equation}
\begin{aligned}
\hat H&=
\sum_{q} \hbar\Delta \hat{r}_q^\dagger\hat{r}_q+\sum_{n,q} \Eblue_{n,q} \hat{b}_{n,q}^\dagger \hat{b}_{n,q}
\\&
+\sum_{n,q} \hbar g_{n,q}\left(\hat{r}_q\hat{b}^\dagger_{n,q}+\text{H.c.}\right)\equiv\sum_{q} \hat H_q
\end{aligned}
\end{equation}
such that the Hamiltonian decouples into independent quasimomenta $[\hat H_q,\hat H_{q'}]=0$, where
\begin{equation}
\begin{aligned}
\hat H_q&=
 \hbar\Delta \hat{r}_q^\dagger\hat{r}_q+\sum_{n} \Eblue_{n,q} \hat{b}_{n,q}^\dagger \hat{b}_{n,q}
\\&
+\sum_{n} \hbar g_{n,q}\left(\hat{r}_q\hat{b}^\dagger_{n,q}+\text{H.c.}\right).
\end{aligned}
\end{equation}
If we take $\hat c_{n,q}^\dagger=\alpha_{n,q}\hat r_q^\dagger+\sum_{n'} \beta_{n,n',q}\hat b^\dagger_{n',q}$ to create a quasiparticle (polariton) in the band $n$ with quasimomentum $q$, then the eigenvalue condition $\hat H_q\hat c_{n,q}^\dagger \ket{0}=\Epurple_{n,q}\hat c_{n,q}^\dagger \ket{0}$ leads to the secular equations \cite{Shi2018_2}
 \begin{equation}
 \left\lbrace
  \begin{array}{l}
(\Epurple_{n,q}-\hbar\Delta)\alpha_{n,q}=\sum_{n'} \hbar g_{n',q}\beta_{n,n',q}
\\
(\Epurple_{n,q}-\Eblue_{n',q})\beta_{n,n',q}=\hbar g_{n',q}\alpha_{n,q}
  \end{array}\right.
 \end{equation}
which allow for the determination of the amplitudes $\alpha_{n,q}$ and $\beta_{n,n',q}$ (which alternatively follow from Eq. (27) in \cite{Lanuza2021_2}) and yield the equations for the polariton dispersion relation
\begin{equation}\label{eq:band structure}
\Epurple_{n,q}-\hbar\Delta=\sum_{n'}\frac{\hbar^2g_{n',q}^2}{\Epurple_{n,q}-\Eblue_{n',q}}.
\end{equation}
as given in the main text.
 
\subparagraph{\it Polariton Bose-Hubbard Hamiltonian.}

Here we show how the band structure and interactions derived above leads to a polariton Bose-Hubbard Hamiltonian. Since the non-interacting Hamiltonian in lattice-momentum space is quadratic and has energies given by \eqref{eq:band structure}, it reduces to
\begin{equation}
    \hat H =\sum_{n,p}\Epurple_{n,q}\hat c^\dagger_{n,p}c_{n,p}.
\end{equation}

Thanks to the independence between quasimomenta, if there is some hopping $J$ for $\ket{r}$ atoms, this can be accounted for by simply making the detuning $q$-dependent, $\Delta\to\Delta_q=\Delta+4J\sin^2\left(\frac{\pi q}{2k}\right)^2$.

Reapplying \eqref{eq:operatorFT} to the $\hat{c}_{n',p}$ operators, one can write the Hamiltonian in position space as
\begin{equation}
 \hat H=-\sum_{n,j,j'} \bar{J}_{n,j-j'}\hat c_{n,j}^\dagger \hat c_{n,j'}
\end{equation}
in terms of the hopping coefficients
\begin{equation}
     \bar{J}_{n,j-j'}=-\sum_q\Epurple_{n,q} e^{i\pi (j-j')q/k}.
 \end{equation}
 
For negative detunings and modest microwave couplings ($\Omega/|\Delta|< 1$) an experimental run that starts with $\ket{r}$ atoms remains dominated by $\ket{r}$, corresponding to a ground polariton band ($n=1$) that is roughly sinusoidal, meaning that we can approximate as if there is only nearest neighbor hopping ($\bar{J}_{1,\pm 1}\equiv\bar{J}$ and $\bar{J}_{1,\lvert n\rvert>1}\approx0$). The constant $-\bar{J}_{1,0}=\sum_q\Epurple_{n,q}$ is the energy of a polariton Wannier function, which is relevant for the calculation of $\bar\omega_0$ in Fig.~\ref{FIG:onsite}. This can be combined with the local energy shift generated by the harmonic confinement of the experiment into a site-dependent  energy $\bar\varepsilon_j$. Omitting the band indices ($\hat c_{1,j}\equiv\hat c_{j}$), the above Hamiltonian then reduces to
\begin{equation}
 \hat H=-\bar J \sum_j(\hat c_{j+1}^\dagger \hat c_{j}+\hat c_{j}^\dagger \hat c_{j+1})+\sum_{j}\bar\varepsilon_j\hat c_{j}^\dagger \hat c_{j},
\end{equation}
which together with the polariton interaction $\hat{H}_\text{int}$ gives the Bose-Hubbard Hamiltonian. The interaction itself is given by
\begin{align}
\hat{H}_\text{int} = \frac{\bar{U}}{2} \sum_j \hat c_{j}^\dagger \hat c_{j}^\dagger \hat c_{j} \hat c_{j}
\end{align}
where we calculate
\begin{align}
\bar{U} &= \frac{1}{2}\mel{0}{\hat c_{0} \hat c_{0} \hat{H}_\text{int} \hat c_{0}^\dagger \hat c_{0}^\dagger}{0}\nonumber\\
& = \bar{g}\left(\int \dif z \abs{\psi_{r}(z)}^4\right. +  \int \dif z \abs{\psi_{b}(z)}^4\nonumber\\
&\quad + 2 \left. \int \dif z \abs{\psi_{r}(z)}^2 \abs{\psi_{b}(z)}^2\right)
\end{align}
in terms of the $\ket{r}$ and $\ket{b}$ components  of the Wannier function, $\psi_{r}(z)=\bra{z,r}\hat{c}_0^\dagger\ket0$ and $\psi_{b}(z)=\bra{z,b}\hat{c}_0^\dagger\ket0$, respectively. Here $\bar{g} =2\hbar^2 a/(m a_{0x}a_{0y})$ depends on the harmonic confinement in the tubes ($a_{0x}$ and $a_{0y}$) and the (small) dependence of the scattering length $a\approx5.3$~nm  on the internal state of the atom is neglected.

\bibliographystyle{apsrev}

\end{document}